\documentstyle[12pt]{article}
\begin{document}
\large

\begin{center}
{\bf ON THE COMPOUND STRUCTURES OF THE NEUTRINO
MASS AND CHARGE}
\end{center}
\vspace{1cm}
\begin{center}
{\bf Rasulkhozha S. Sharafiddinov}
\end{center}
\vspace{1cm}
\begin{center}
{\bf Institute of Nuclear Physics, Uzbekistan Academy of Sciences,
Tashkent, 702132 Ulugbek, Uzbekistan}
\end{center}
\vspace{1cm}

The mass and charge of a particle correspond to the most diverse form
of the same regularity of the nature of this field. As a consequence,
each of all possible types of charges testifies in favor of the existence
of a kind of the inertial mass. Therefore, to investigate these features we
have established the compound structures of mass and charge. They can explain
also the availability of fundamental differences in the masses as well as
in the charges of Dirac and Majorana neutrinos.

\newpage
One of sharply expressed features of the interaction of Dirac
$(\nu_{D}\neq \overline{\nu}_{D})$ and Majorana $(\nu_{M}=\overline{\nu}_{M})$
neutrinos with field of emission \cite{1,2,3} is the connection between these
phenomena and character of the structure of fermions \cite{4,5,6,7}
themselves. At the same time a question about the nature, similarity and
difference of masses of neutrinos of both types \cite{8} remains thus far
not finally investigated.

The nature is, according to well known considerations, created so that all
the forces in it be have the unified regularity. From this point of view,
becomes possible use the Newton law of gravity as the Coulomb law and vice
versa. In other words, these forces correspond to the most diverse form of
the same action. Exactly the same one can as the interacting objects choose
the two of neutrinos. Such a procedure, however, takes place regardless of
the neutrino structure of whether it is the Dirac or the Majorana fermion.
In this a hard connection is said between the inertial mass of a particle
and its physical nature.

Our study of elastic scattering of electrons and their neutrinos on a
spinless nucleus shows \cite{9,10} clearly that if the neutrino is the
four - component particle $(\nu=\nu_{D}=\nu_{e})$ having a Dirac mass
$m_{\nu_{D}},$ it must possess both normal and anomalous electric charges.
According to these data, the neutrino full electric charge in the static
limit has the size
\begin{equation}
e_{\nu}=-\frac{3eG_{F}m_{\nu}^{2}}{4\pi^{2}\sqrt{2}}, \, \, \, \, e=|e|.
\label{1}
\end{equation}

Such a picture leading to the flip of the neutrino spin \cite{11} and
reflects the fact that the mass and charge of a particle correspond to
two form of the same regularity of the nature of its structure \cite{9,10}.

For further purposes of a given work it is desirable to remind about the
electron mass. From point of view of the classical theory of electromagnetic
mass \cite{12}, the availability of the eigenenergy $E_{0}$ of the electron
electrostatic field implies the existence of the electric part of the
electron rest mass:
$$m_{e}^{em}=\frac{E_{0}}{c^{2}}.$$

The assumption has even been speaked out that all the mass of the electron
is equal to its electromagnetic mass. This idea was simply called a hypothesis
of field mass.

From our earlier developments, we find that a particle mass is strictly
multicomponent. One of them corresponds to the electric charge and can be
called a Coulomb mass. Insofar as the electrically neutral neutrino \cite{13}
is concerned, its mass does not contain the part, at which it would have as
well as an electric charge.

The difference in the masses of Dirac $(\nu=\nu_{D})$ and Majorana
$(\nu=\nu_{M})$ neutrinos is observed because each of all possible
types of charges of the same neutrino arises as a consequence of the
availability of a kind of the inertial mass. Thereby this mechanism leads
to the appearance of the united rest mass $m_{\nu}^{U}$ of the neutrino equal
to all the mass of a given particle. Its general structure at the account of
nonweak \cite{14,15} and unknown properties of the neutrino has the form
\begin{equation}
m_{\nu}^{U}=m_{\nu}^{E}+m_{\nu}^{W}+m_{\nu}^{S}+...,
\label{2}
\end{equation}
where $m_{\nu}^{E}, \, \, \, \, m_{\nu}^{W}$ and $m_{\nu}^{S}$ denote
respectively the electric, weak and strong components of mass. Such a sight
to the origination of $m_{\nu}^{U}$ quality explains the presence of the
united charge $e_{\nu}^{U}$ of the neutrino equal to all the charge of a
given fermion which contains the electric $e_{\nu}^{E},$ weak $e_{\nu}^{W},$
strong $e_{\nu}^{S}$ and some other parts:
\begin{equation}
e_{\nu}^{U}=e_{\nu}^{E}+e_{\nu}^{W}+e_{\nu}^{S}+....
\label{3}
\end{equation}

So, it is seen that the mass and charge of a Dirac particle include in self
both electric and unelectric components. Of course, our formula (\ref{1})
characteize only a Coulomb mass dependence of the electric charge:
$$m_{\nu}=m_{\nu_{D}}^{E}, \, \, \, \, e_{\nu}=e_{\nu_{D}}^{E}.$$

Using this and by following the fact that the force of Newton attraction
between the two neutrinos is less than the force of their Coulomb repulsion,
we find that
\begin{equation}
m_{\nu_{D}}^{E}> 1.53\cdot 10^{-3}\ {\rm eV},
\label{4}
\end{equation}
\begin{equation}
e_{\nu_{D}}^{E}> 1.46\cdot 10^{-30}\ {\rm e}.
\label{5}
\end{equation}

It is clear, however, that the finding values of (\ref{4}) and (\ref{5}) are
incompatible with the available laboratory data \cite{16}. At the same time
this circumstance may serve as some confirmation of the availability of
compound structures of mass and charge.

Thus, it follows that if neutrinos are of electrically neutral
$(\nu=\nu_{M})$ then $m_{\nu_{M}}^{E}=0, \, \, \, \, e_{\nu_{M}}^{E}=0,$
and the size of $m_{\nu_{M}}^{U}$ and $e_{\nu_{M}}^{U}$ are reduced
to the form
\begin{equation}
m_{\nu_{M}}^{U}=m_{\nu_{M}}^{W}+m_{\nu_{M}}^{S}+...,
\label{6}
\end{equation}
\begin{equation}
e_{\nu_{M}}^{U}=e_{\nu_{M}}^{W}+e_{\nu_{M}}^{S}+....
\label{7}
\end{equation}

Comparing their with (\ref{2}) and (\ref{3}) at $\nu=\nu_{D},$ it is easy
to observe the fundamental differences in the masses as well as
in the charges of neutrinos of the different nature.

Of course, the above - noted regularities of general picture of massive
neutrinos extending well known hypothesis of field mass meet with many
problems which require the study of the structure and property of each
of existing types of charges and masses.

\begin{center}
{\bf Acknowledgement}
\end{center}

I would like to express my gratitude to Professor Burkhan K. Kerimov for
his numerous valuable discussions and critical remarks that was useful in
improving the manuscripit.

\end{document}